\documentclass[twocolumn, superscriptaddress,  floatfix, amssymb,unsortedaddress, aps, pre, amssymb,amsmath,color,showpacs]{revtex4}

\usepackage{graphics}

\newcommand {\rsq}{\langle R^2 \rangle}

\begin{document}

\title{Topological versus rheological entanglement length in primitive path analysis protocols}

\author{Ralf Everaers}
\affiliation{
Universit\'e de Lyon, France;
CNRS, UMR 5672;
ENS de Lyon, Laboratoire de Physique \& Centre Blaise Pascal, 46, all\'ee d'Italie, Lyon, F-69364, France}

\date{\today}

\begin{abstract}  
Primitive path analysis algorithms are now routinely employed to analyze entanglements in computer simulations of polymeric systems, but different analysis protocols result in different estimates of the entanglement length, $N_e$. Here we argue that standard {\em PPA} measures the {\em rheological} entanglement length, $N_e^{rheo}$, typically employed by tube models and relevant to quantitative comparisons with experiment, while codes like {\em Z} or {\em CReTA} also determine the {\em topological} entanglement length, $N_e^{topo}$. For loosely entangled systems, a simple analogy between between phantom networks and the mesh of entangled primitive paths suggests $ N_e^{rheo}\approx 2 N_e^{topo}$. This result is in excellent agreement with reported values for poly-ethylene, poly-butadiene and bead-spring polymer melts.
\end{abstract}


\maketitle

Understanding the role of entanglements in the dynamics of high molecular weight polymeric 
liquids is a classical subject of polymer physics~\cite{doi86, mcleish02}.
Tube models \cite{edwards67,doi86, mcleish02} describe the neighborhood of a single chain in terms of a mean-field confinement potential, which transiently localizes thermally
fluctuating chains in a tube-like region in space. 
Primitive paths were originally introduced in a thought experiment to determine
the tubes confining individual polymers in an entangled polymer melt or
network.  The idea is to identify the random walk-like tube axis with the
shortest (``primitive'') paths between the end points of the original chains into
which the chain contours can be contracted without crossing each
other.  
Corresponding primitive path {\em analysis} (PPA) algorithms~\cite{PPA,PPA2,Kroeger_cpc_05,ShanbhagLarson_prl_05,ZhouLarsonMM2005,Tzoumanekas_mm_06,Uchida_jcp_08} have become a standard tool in computer simulations of polymeric systems.
In particular, we have claimed that the primitive path analysis  
 (i) endows the tube models with predictive power for dynamical processes in flexible chain melts~\cite{HouSGE_prl_10} and (ii) provides a {\em quantitative}
unified picture of chain entanglement over a much wider class of polymeric systems~\cite{Uchida_jcp_08}. However,  estimates of the melt entanglement lengths based on different PPA variants seem to vary within a factor of two~\cite{HoyPRE09}. Is there something amiss?

The present note was triggered by a recent study by Hoy {\it et al.}~\cite{HoyPRE09}, which contains a comparison of PPA and Z results for polyethylene and bead-spring polymer melts. The original PPA algorithm~\cite{PPA,PPA2} replaces the intra-chain interactions by harmonic bonds between neighboring beads. In the course of an energy minimization with fixed chains ends, the bond springs try to reduce the bond length to zero and pull the chains taut. The inter-chain excluded volume interactions ensure that different chains do not cut through each other. In contrast, Kr\"oger's numerically much  more efficient Z-code uses geometrical operations to shorten primitive paths, which are treated as tensionless lines rather than multibead chains. Other PPA algorithms~\cite{ShanbhagLarson_prl_05,ZhouLarsonMM2005,Tzoumanekas_mm_06,Uchida_jcp_08} follow similar recipes. 

A first observation by Hoy {\it et al.}~\cite{HoyPRE09} well worth repeating is that there is {\em no} substantial difference between the results from the PPA and the Z algorithms, if one compares {\em identical} observables (typically the primitive path length, $\langle L_{pp} \rangle$ or $\sqrt{\langle L_{pp}^2 \rangle}$). Nevertheless, their results suggest a relation of $N_e^{PPA}\approx 2 N_e^{Z}$ between entanglements lengths inferred following the PPA and Z protocols: PPA practitioners measure the number of monomers per PP Kuhn length, $a_{pp}$, while $N_e^{Z}$ is the number of monomers between interior kinks at positions, where two primitive paths intersect. That $N_e^{Z}<N_e^{PPA}$ and that there is hence an orientational correlation between primitive path segments was first pointed out by Tzoumanekas and Theodorou~\cite{Tzoumanekas_mm_06}. For loosely entangled systems, the results should agree on a scaling level~\cite{Uchida_jcp_08}. But why is there a factor of two? And are we sure, that it is really $N_e^{PPA}$ and not $N_e^{Z}$ which should be compared to experimentally measured entanglement lengths?

In the following, we will argue that these questions are easily resolved by considering a corresponding problem in the framework of the classical theory of rubber elasticity~\cite{treloar75}, which deals with the behavior of networks of Gaussian chains under strain. In particular, we will find that the present issue is related to the ``frontfactor'' debate~\cite{JamesGu_jcp_47,Duiser_conf,Graessley_mm_75,Flory76} for the predicted elastic moduli.
The junction affine model~\cite{kuhn36} and the phantom model~\cite{JamesGu_jcp_43,JamesGu_jcp_47} both predict shear moduli of $G=\gamma \rho_{strand} k_BT$, where $\rho_{strand}=\rho/N_s$ denote the number density of elastically active network strands and monomers respectively. But while $\gamma \equiv1$ for the junction affine model, the phantom model predicts a dependence $\gamma= (1-2/f)$ on the functionality of the junction points (or, more appropriately, the cycle rank~\cite{Flory76} of the network). It took until 1976 for Flory~\cite{Flory76} to resolve the discussion. The salient points are apparent from the following simplified argument~\cite{everaers98}:  in a phantom network, the thermal fluctuation of a junction point around its rest position are of the order of $N_s b^2/f$, where $\rsq=N_s b^2$ is the mean-square end-to-end distance of the network strands of length $N_s$ in a corresponding linear chain melt. If one assumes that neighboring junction points fluctuate independently (obviously a simplification), then the {\em thermal} fluctuations of the extension of a network strand are given by $2/f \rsq$. Since instantaneous random crosslinking does not change the static {\em ensemble} average of the strand end-to-end distances, one can thus deduce the {\em quenched} part of the fluctuations:   the mean-square {\em average} strand extension in the network is $(1-2/f) \rsq \equiv \gamma \rsq$. The same prefactor appears in the shear modulus predicted by the phantom model, 
\begin{equation}\label{eq:G_phantom}
G_{phantom} = \frac{\rho\, k_BT }{N_s \,/ \left(1-2/f \right)}
\end{equation}
because the thermal fluctuations of a phantom network are invariant under strain, while the  {\em average} strand extensions undergo an affine transformation with the macroscopic deformation~\cite{JamesGu_jcp_47}. The earlier junction affine model neglects this subtle mathematical point and derives the elastic reponse from an affine transformation of the distribution of instantaneous end-to-end distances, so that $\gamma\equiv1$. In real networks, the situation is intermediate between these two limits, since entanglements contribute to the junction localization~\cite{ronca75,Flory_jcp_77,everaers99}. 

For a phantom network prevented from collapse by periodic boundary conditions or surface anchored junction points, PPA converges to the ground state with all junction points localized at their average positions~\cite{PermSet3}. To understand the relation between the above consideration and the interpretation of PPA results, we need to analyze the statistics of (primitive) paths through the network (these paths could, for example, correspond to long, multiply crosslinked precursor chains~\cite{SGE_prl_04,SGE_poly_05}) . On large scales, $N\gg N_s$, the paths follow the same random walk statistics as the original chains with $\rsq(N)= N b^2$. In contrast, PPA smoothens out local thermal fluctuations on the mesh scale, so that the paths become locally straight between junction points. By construction, the Kuhn length, $a_{pp}$, of the paths is given by 
\begin{equation}
a_{pp} = \frac{\rsq(N)}{L_{pp}(N)} = \frac{b N_s}{\sqrt{(1-2/f) N_s}}
\end{equation}
where $L_{pp}=\frac{N}{N_s} \sqrt{(1-2/f) \rsq(N_s)}$ is the contour length distance along the path and where we have deduced the mean-square contour length per network {\em strand} from the above arguments for phantom networks. In analogy to the standard PPA protocol, we now consider the number, $N_s^{PPA}$, of monomers per Kuhn length of the primitive path:
\begin{equation}
N_s^{PPA} 
   = \frac{a_{pp}}{L_{pp}/N} 
   = \frac{N_s}{1-2/f} 
\end{equation}
In particular, we see by comparison with Eq.~(\ref{eq:G_phantom}) that 
\begin{equation}\label{eq:G_PPA}
G=\frac{\rho\, k_BT}{N_s^{PPA}}.
\end{equation}

Neglecting the subtle difference between slip-links~\cite{GraessleyPear_jcp_77} and crosslinks, we see that the number of monomers between entanglements or {\em topological entanglement length}, $N_e^{topo}=N_e^Z$  is the direct analogue of the strand length $N_s$ in a polymer network. Nevertheless, it is the less intuitive  {\em rheological entanglement length}, $N_e^{rheo}=N_e^{PPA}$, which should be compared to experimentally determined values of $N_e$ , because the latter are derived from the melt plateau modulus (or the closely related entanglement modulus~\cite{mcleish02}) and the relation  $G_e = \rho\, k_BT/N_e$.  

The reported values of $ N_e^{rheo}/N_e^{topo}=2.6 $~\cite{Tzoumanekas_mm_06} and $ N_e^{rheo}/N_e^{topo}=1.9$ and  2.0 \cite{HoyPRE09} for poly-ethylene, $ N_e^{rheo}/N_e^{topo}=2.1 $ \cite{Tzoumanekas_mm_06} for poly-butadiene and $ N_e^{rheo}/N_e^{topo}=1.8 $  \cite{HoyPRE09} for Kremer-Grest melts~\cite{Kremer90} are
compatible with the interpretation of entanglements as binary contacts (or $f=4$ functional sliplinks) between primitive paths. We emphasize, that the present analogy only holds for fully flexible, loosely entangled chains.  Tightly entangled chains fall into a different regime with $ N_e^{rheo}\gg N_e^{topo}$ ~\cite{Uchida_jcp_08}.


\end{document}